%% file: main.tex
\documentclass[lettersize,journal]{IEEEtran}
\usepackage{amsmath,amsfonts}
\usepackage{array}
\usepackage{textcomp}
\usepackage{stfloats}
\usepackage{url}
\usepackage{verbatim}
\usepackage{graphicx}
\usepackage{cite}
\usepackage{color}
\usepackage{amsmath,amssymb,amsthm,mathrsfs}
\usepackage{algpseudocode} 
\usepackage[ruled,norelsize,linesnumbered]{algorithm2e}
\usepackage{subcaption}
\usepackage{todonotes}
\usepackage{bm}
\makeatletter
\newcommand{\removelatexerror}{\let\@latex@error\@gobble}
\newcommand{\nosemic}{\renewcommand{\@endalgocfline}{\relax}}% Drop semi-colon ;
\newcommand{\dosemic}{\renewcommand{\@endalgocfline}{\algocf@endline}}% Reinstate semi-colon ;
\makeatother

\newcommand*{\note}[1]{\textcolor{red}{#1}}
\newcommand*{\cpy}[1]{\textcolor{blue}{#1}}
\newcommand{\algname}{{CC-MASAC}\xspace}
\newcommand{\Algtwo}{PSRU\xspace}

\begin{document}

\title{Unleashing Collaborative Computing for Adaptive Video Streaming with Multi-objective Optimization in Satellite Terrestrial Networks}

%\title{Unleashing the Collaborative Computing for Adaptive Video Streaming with QoE Optimization in Satellite Terrestrial Networks \note{[?]}}

\author{\IEEEauthorblockN{Zhishu Shen\textbf{\textsuperscript{*}},~\IEEEmembership{Member,~IEEE,} 
Qiushi Zheng\textbf{\textsuperscript{*}},~\IEEEmembership{Member,~IEEE}\thanks{\textbf{*} Both authors contributed equally to this work. (Corresponding author: Zhishu Shen)},
Ziqi Rong,
Jiong~Jin,~\IEEEmembership{Member,~IEEE,}
Atsushi~Tagami,~\IEEEmembership{Member,~IEEE,}
and Wei~Xiang,~\IEEEmembership{Senior Member,~IEEE}}

\thanks{Zhishu Shen and Ziqi Rong are with the School of Computer Science and Artificial Intelligence, Wuhan University of Technology, Wuhan, China (e-mail: z\_shen@ieee.org, rongziqi@whut.edu.cn).}
\thanks{Qiushi Zheng and Jiong Jin are with the School of Science, Computing and Engineering Technologies, Swinburne University of Technology, Melbourne, Australia (e-mail: \{qiushizheng, jiongjin\}@swin.edu.au).}
\thanks{Atsushi Tagami is with KDDI Research, Inc., Japan (e-mail: at-tagami@kddi.com).}
\thanks{Wei Xiang is with the School of Computing, Engineering and Mathematical Sciences, La Trobe University, Melbourne, Australia (e-mail: w.xiang@latrobe.edu.au).}
% <-this % stops a space
%\thanks{Zhishu Shen is the corresponding author.}
%\thanks{Manuscript received March X, 2024.}
}

\markboth{Journal of \LaTeX\ Class Files,~Vol.~14, No.~8, August~2021}%
{Shell \MakeLowercase{\textit{et al.}}: A Sample Article Using IEEEtran.cls for IEEE Journals}


\maketitle

\begin{abstract}
Satellite-terrestrial networks (STNs) are anticipated to deliver seamless IoT services across expansive regions. Given the constrained resources available for offloading computationally intensive tasks like video streaming, it is crucial to establish collaborative computing among diverse components within STNs. In this paper, we present the task offloading challenge as a multi-objective optimization problem, leveraging the collaboration between ground devices/users and satellites. We propose a collaborative computing scheme that optimally assigns computing tasks to various nodes within STNs to enhance service performance including quality of experience (QoE). This algorithm initially dynamically selects an end-to-end path that balances service time and resource utilization. For each selected path, a multi-agent soft actor-critic (MA-SAC)-based algorithm is introduced to make adaptive decisions and collaboratively assign optimal heterogeneous resources to the given computing tasks. In this algorithm, the ground station bridging satellite network and terrestrial network is treated as agent to extract the information from both STNs and users. Through MA-SAC, multiple agents cooperate to determine the adaptive bitrate and network resources for the arriving tasks. The numerical results demonstrate that our proposal outperforms comparative schemes across various computing tasks in terms of various criteria. 

\end{abstract}


\input{1_introduction}

\input{2_relatedwork}

\input{3_model}

\input{4_problem}

\input{5_algorithm}

\input{6_experiment}

\input{7_conclusion}

\bibliographystyle{IEEEtran}
\bibliography{ref}

\end{document}

%% file: 1_introduction.tex
\section{Introduction}
\IEEEPARstart{T}{he} promising development of Internet of Things (IoT) presents abundant opportunities to realize intelligent data services, including video streaming analytics for wide applications, such as smart transportation~\cite{Wu2022ASO}, smart factory~\cite{QiuCST20}, and smart building~\cite{ShenIEM21}. On the other hand, the development of intelligent applications in remote or rural areas has been stagnant due to the high cost and limited returns associated with establishing traditional terrestrial networks to serve sparsely populated and clustered remote or rural areas~\cite{YaacoubIEEE20}. The remote or rural areas herein refer to those where network communication infrastructure and its associated functions such as data storage, electricity supply and system maintenance are unreliable or even unavailable. Therefore, new computing solutions capable of facilitating high-quality video streaming services are essential for these remote or rural areas.

Low earth orbit (LEO) satellites are regarded as a promising solution to extend the connectivity beyond current communication networks~\cite{CentenaroCST21,KodheliCST21,GuoCST22}. The progress in satellite manufacturing and launching technologies has advanced the deployment of LEO satellites in space. The integration of terrestrial networks and satellite networks in satellite-terrestrial networks (STNs) aims to provide ubiquitous network services, ensuring widespread and reliable connectivity for global users~\cite{VaeziCST22}. Additionally, the potential benefits of integrating computing capabilities into satellites have been extensively explored as a means to decrease data processing and communication latency in STNs~\cite{ChenWC22,HayderCST23,shen2023survey}. Recently, there have been advancements in the development of satellite computing platforms such as Tiansuan Constellation~\cite{WangIoT23} and Orbital Edge Computing~\cite{DenbyASPLOS23}. These platforms are anticipated to facilitate real-time data processing and reduce the volume of data that needs transmission through STNs, thereby lowering associated costs.

%Although satellites can provide seamless connectivity to massive devices/users located in wide areas including remote/rural areas, the STNs tend to have higher communication delay due to the long propagation distance of satellite-terrestrial links~\cite{ZhuJSAC23}. 

In consideration of the limited resources available to end-users on the ground, it is essential to offload computing tasks to the computation server on satellites to improve service quality. However, as the volume and variety of data generated at the ground site increase, the available resources from both ground nodes and satellites become more depleted. This depletion can lead to service delays or even interruptions, resulting in a decrease in user quality of experience (QoE). Particularly, it is critical for a task offloading scheme to address the emerging video streaming services that process substantial amounts of streaming data. The optimization problem of task offloading in STNs, which involves selecting appropriate ground nodes and satellites for each computing task while meeting service requirements, is computationally intensive~\cite{TangIoT21,ChenTMC24}. To mitigate computational complexity, various approaches have been proposed such as the alternating
direction method of multipliers (ADMM)~\cite{AliaTMC24}, particle swarm optimization (PSO)~\cite{ZhuJSAC23}, and deep deterministic policy gradient (DDPG)~\cite{ZhangIoT23}. Meanwhile, adaptive bitrate video streaming stands out as a prominent solution for enhancing user QoE in video streaming  services~\cite{JiangCST21}. In this solution, the video is divided into a sequence of segments, each encoded at different bitrates, allowing flexible video quality selection to support task offloading in STNs. In order to meet the high QoE standards for video streaming services, it is crucial for STNs to dynamically adjust the bitrate of each segment based on the current network conditions.

%In the process of designing computing task offloading and adaptive bitrate control for video streaming services to ensure QoE within STNs, several challenges remain:

The challenges of designing computing task offloading and adaptive bitrate control for video streaming services to ensure QoE within STNs are as below:

\begin{enumerate}
    \item How to solve the task offloading optimization problem within dynamic network environments in STNs? (Section~\ref{subsec:framework}) 
    \item How to coordinate heterogeneous nodes in STNs to efficiently utilize the limited computational and communication resources provided by STNs? (Section~\ref{subsec:path_selection} and \ref{subsec:masac}) 
    \item How to meet the requirements of multiple criteria from the perspective of both user and network operation for different types of computing tasks? (Section~\ref{subsec:masac})
\end{enumerate}  

In this article, we propose a collaborative satellite computing scheme for adaptive video streaming in STNs. This scheme is designed to solve the multi-objective problem that maximizes the QoE and heterogeneous resource utilization while minimizing the service delay. Since this optimization problem is NP-hard, we decompose the original problem into two sub-problems: path selection and resource allocation with adaptive bitrate control. 
It first obtains the current optimal path candidate and then decides the bitrate for the given task with heterogeneous resource allocation using multi-agent soft actor-critic (MA-SAC). The ground station, acting as the agent, connects the satellite network and terrestrial network, analyzing information from both the network and user sides. Enabled by multiple agents, the developed MA-SAC algorithm strives to attain the optimal policy for task offloading with adaptive video streaming by extracting the dynamic information STNs to maximize the entropy. Consequently, our scheme is capable of accommodating massive computing tasks while endeavoring to meet various service requirements. The main contributions of this work are outlined as follows:\begin{itemize}
    \item  \textbf{Problem Formulation}: A multi-objective task offloading optimization problem is modeled for STNs to investigate the interaction mechanism between numerous ground devices and the network. This problem encompasses two distinct computing tasks with varying demands on network communication and computation. Based on the system model considering the unique characteristics of ground nodes and satellites, we define QoE, service delay, energy consumption and heterogeneous resource usage as the objective function to fulfil the requirements of various services. 
    %\item  \A dynamic LEO constellation network framework is designed for IoT services in infrastructure-less environments. In this framework, a virtual digital twins network is utilized to map the physical networks for dynamic update on network's heterogeneous resource states for both terrestrial network and satellite network. 
    \item \textbf{Algorithm}: To mitigate computational complexity, the original optimization problem is decomposed into two sub-problems: path-selection and resource allocation with adaptive bitrate control. For each arriving computing task, the path selection algorithm dynamically selects an end-to-end path on a given score, considering the service delay and resource utilization. The second sub-problem is the bitrate adaption optimization for improving QoE and reducing energy consumption during the task offloading process in STNs. Leveraging the selected path candidate, the resource allocation algorithm employs MA-SAC for heterogeneous resource allocation while determining the optimal bitrate for data transmission. %To the best of the author's knowledge, this is the first work that introduces the bitrate adaption optimization for improving QoE and reducing energy consumption during the task offloading process in STNs. 
    \item \textbf{Evaluation}: Extensive experiments are conducted to evaluate the performance of our proposed scheme across various types of computing tasks in STNs. The results obtained from comprehensive experiments demonstrate that our proposal outperforms several comparative schemes. Specifically, our proposed scheme can significantly improve the task completion rate, QoE and service delay performance while maintaining low energy consumption.%, \note{XXX}\% reduction in service delay, and \note{XXX}\% improvement in computational resource usage. 
\end{itemize}

The remainder of this paper is organized as follows: Section~\ref{sec:relatedwork} summarizes the related work. Section~\ref{sec:model} presents the system model, followed by the problem formulation in Section~\ref{sec:formulation}. Section~\ref{sec:algorithm} presents our proposed algorithm that realizes collaborative computing for adaptive video streaming in STNs. Section~\ref{sec:evaluation} demonstrates the evaluation results of the proposed scheme against the comparable ones. Section~\ref{sec:conclusion} summarizes this work. The main notations used in this paper are summarised in Table  \ref{tab:notation}.

%% file: 2_relatedwork.tex
\section{Related Work}~\label{sec:relatedwork}
%\subsection{Satellite computing\note{[?]}}

%\subsection{STNs architecture for video streaming analytics \note{[?]}}

\subsection{Task offloading for STNs}
In the research on task offloading for STNs comprising a single satellite, Ding \textit{et al.} investigated the effectiveness of introducing a high altitude platform station (HAPS) and a LEO satellite with computing capability to the terrestrial networks. For this purpose, a communication and computation resource allocation problem is formulated that minimizes energy consumption~\cite{DingTWC22}. Zhu \textit{et al.} proposed a particle swarm optimization (PSO)-based approach that cooperates with multiple local devices, a satellite (edge server) and a cloud server on the ground to optimize service delay in STNs~\cite{ZhuJSAC23}. 
Regarding the studies focusing on task offloading in multi-satellite networks, 
Chen \textit{et al.} proposed a double deep Q-learning-based approach to optimize energy consumption and latency for task offloading~\cite{ChenIoT22}. Wang \textit{et al.} used Stackelbery mean-field game to formulate the interactive problem between ground users and ultra-dense LEO satellites~\cite{WangJSTSP23}. 

Among the state-of-the-art literature, deep reinforcement learning (DRL)-based algorithms have recently garnered attention~\cite{QiuTVT19, DaiTMC23, ChenIoT22, ZhangIoT23}. DRL leverages deep neural networks (DNN) to optimize the policies and actions, which is essential to achieve scalable optimization for high-dimensional optimization problems. Considering the complexity of the multi-objective optimization problem in STNs, it remains significant to study how to utilize DRL to squeeze the limited heterogeneous resources provided by STNs for collaborative computing.

\subsection{Adaptive bitrate streaming in STNs}
Multi-access edge computing (MEC)-based adaptive video streaming schemes have already been extensively explored in terrestrial networks. Bayhan \textit{et al.} introduced a network control mechanism to ensure high video quality by utilizing edge-cached video content for functions such as cache assignment and adaptive bitrate control~\cite{BayhanTNSM21}. Huang \textit{et al.} proposed a neural meta-reinforcement learning adaptive bitrate system to rapidly adapt to the specific network conditions including those in 4G/5G wireless networks~\cite{HuangJSAC22}. Wang \textit{et al.} proposed an energy-aware adaptive video streaming scheme that includes a soft
actor-critic (SAC)-based joint uplink transmission and edge transcoding algorithm that maximizes the user QoE~\cite{WangJSAC22}. Qian \textit{et al.} developed a gradient-based video analytic algorithm offloading algorithm that utilizes the edge-cloud collaboration. The objective of this algorithm is to minimize the processing latency of each video stream to achieve real-time video analytics ~\cite{QianTC23}.

In the realm of satellite networks, the existing research primarily focuses on TV broadcast satellite~\cite{AbdulmajeedTVT22}. In contrast to satellite TV services that deliver content from providers to subscribers through TV operators in a broadcast fashion, video streaming services establish direct connections between video sources and users, offering a broader range of service options. Lin \textit{et al.} introduced an adaptive video streaming scheme based on policy proximal optimization (PPO) for LEO satellite constellation to maximize the user QoE~\cite{LinICC23}. Zhao \textit{et al.} presented a contextual multi-armed bandit algorithm to address the multipath adaptive video streaming problems to enhance QoE~\cite{ZhaoGlobecom23}. Optimizing user QoE in video streaming necessitates an efficient solution to handle diverse service types with different service requirements.

%~\cite{XuTPDS24}

%% file: 3_model.tex
\section{System Model}\label{sec:model}
%\subsection{Satellite computing}

\subsection{Network model}\label{subsec:sysmodel}

\begin{figure}[tb!]
    \centering
    \includegraphics[width = \columnwidth,trim=50 0 40 0,clip]{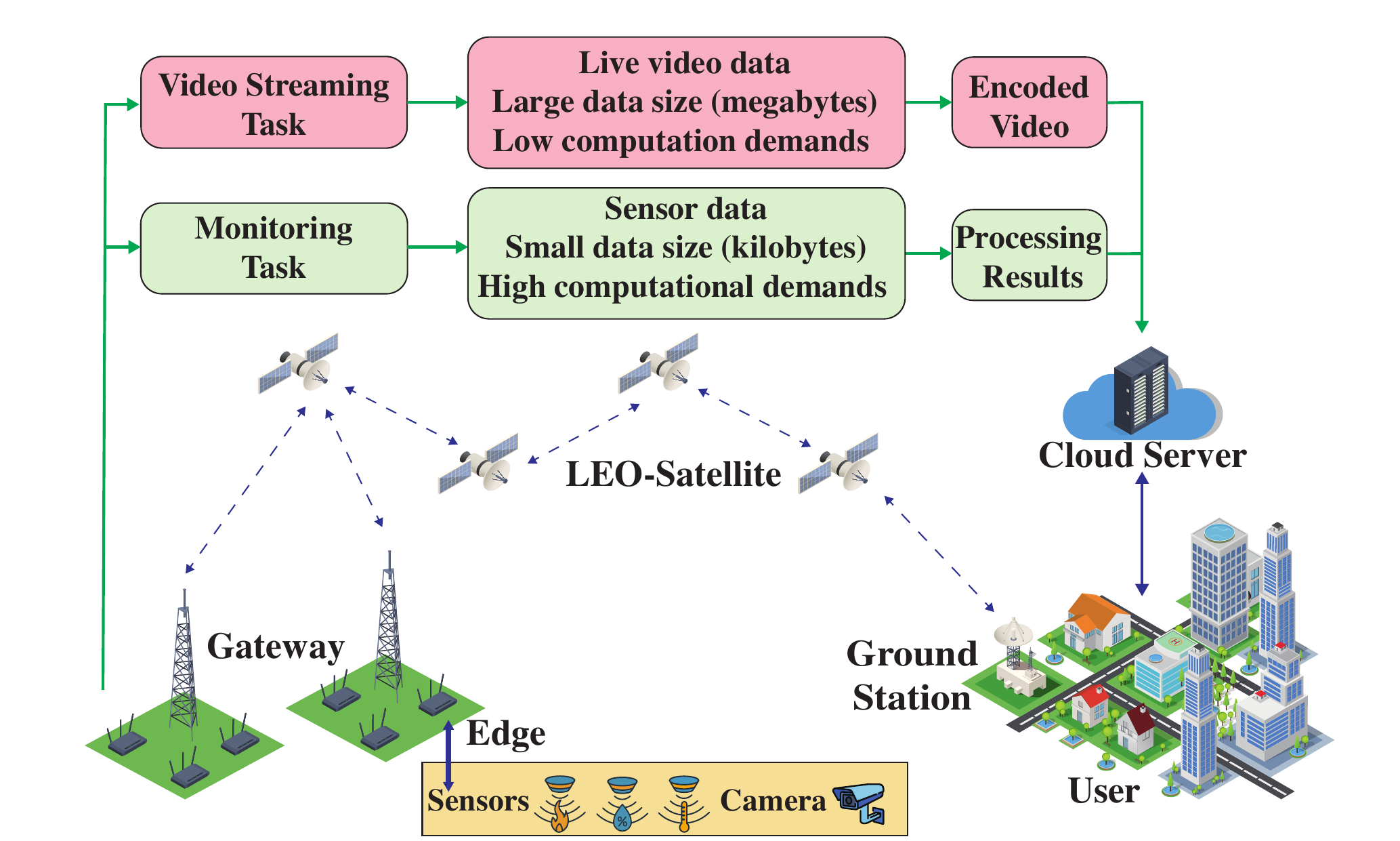}
    \caption{Network model.}
    \label{fig:model}
\end{figure}

\figurename~\ref{fig:model} illustrates the network architecture that provides various services via satellite networks. Within each area, the IoT \textbf{devices} $d \in \mathcal{D}$ generate the computation task, such as the temperature, humidity or farm animal cognition video data processing for the given services. The \textbf{edge} $e \in \mathcal{E}$ on the ground aggregates and processes the data from its managed IoT devices. Meanwhile, the neighbor edges can assist in data processing. If all edges in an area are unable to handle a task, it will be transferred to a regional \textbf{gateway} $g$ which acts as a bridge between satellite network and terrestrial network, i.e., a gateway performs the path selection for both networks and connects them for task offloading. It is important to note that gateways located in distant areas cannot connect to others due to limited communication capabilities in remote or rural regions~\cite{YaacoubIEEE20}.

\begin{table}[tb]

 \caption{Main notations used in this paper.}
 \label{tab:notation}
 \centering
\begin{tabular}{ll}
   \hline
     & Notations \\
   \hline \hline

   % from 。。。to ...
   $B_g$ &  Channel bandwidth between gateway $g$ and edge $e$\\
   $g_{g,e}^{t}$ &   Channel gain from gateway $g$ to edge $e$ \\
   $p_{g,e}$  &   Transmission power from gateway $g$ to edge $e$\\
   $N_0$ &  Noise power spectral density \\  
   $r_{g,e}^\textit{E2G}$ &  Uplink data rate from edge $e$ to gateway $g$ \\
   % 常数不用写
    $r^\textit{E2E}_{e,l}$ &  Maximum data rate from edge $e$ to edge $l$\\
    $B_\textit{s}$ &  Channel bandwidth between gateway $g$ and satellite $i$\\
    
    $r_{g,i}^\textit{G2S}$ &  Uplink data rate from gateway $g$ to the satellite i   \\
    $ r_{k,i}^\textit{S2S}$ & Maximum data rate between satellite $k$ and $i$ \\
    $P_t$ & Power of transmitter \\
    $G_k(i)$ & Average antenna gain of satellite $k$ towards satellite $i$\\
    $L(k,i)$ & Free space path loss between satellite k and i\\
    % 后面加一点
    $f_c$ & Carrier frequency \\
     $\textit{dist}(k,i)$ & Euclidean distance between satellite $k$ and $i$\\ 
     $m$ & Computation task \\
    $D_m$ & Data size of task $m$ \\
    $C_m$ & Required computation resource (CPU cycle) of task $m$\\
    $T^\textit{max}_m$ & Time limitation for task processing of task $m$ \\
    $\alpha_{m,e} $ & Proportion of data processed at the edge \\
    $d_m^\textit{re,LC}$ &  Data size that remains after edge processing  \\
    $c_m^\textit{re,LC}$ &  Remaining computing resources required after edge processing \\
    $\beta_{m,i}$ & Proportion of data processed at satellite\\
    $d_m^\textit{re,SC}$ & Data size that remains after satellite computing \\
    $e^e$ &  Encoding energy consumption for task $m$ \\
    $r^e$ & Encoding bitrate for task $m$ \\
    $\kappa_v$ & Energy efficiency parameter\\
    $e^u$ & Transmission energy consumption for task $m$\\
    $t^u$ &  Uploading time for task $m$\\
    $f$ & CPU frequency \\ 
    $t^{tc}$ & Transcoding time for task $m$ \\
    $t^{LC}_\textit{tran}$ & Transmission delay at edge\\
    $t^{LC}_\textit{prop}$ & Propagation delay at edge \\
    $t^\textit{comp,LC}_{m,k}$ & Computation delay at edge $k$ for task $m$\\
    $C^\textit{comp,LC}_k$ & Computation capacity of edge $k$ \\
    $f^\textit{SAT}_{m,i}$ & Computation capacity of satellite $i$ allocated to current task $m$ \\

    % $B_s$和$B_g$是一个意思吗？
    % 还有p_e 和 p_g_e 是一个意思吗？

   \hline
  \end{tabular}
\end{table}

To realize a regional coverage for data communication, we assume an STN $\mathcal{G}$ = ($\mathcal{V}$: node, $\mathcal{L}$: link), where the node can be an edge, a gateway, or a satellite. \textbf{Satellite} $i\in \mathcal{S}$ empowered by satellite computing capability can process the data received from the gateways. In this paper, we consider two types of video tasks: \textbf{monitoring task} and \textbf{video streaming task}. The monitoring task involves collecting sensor data for scenarios such as wildlife or remote building structural monitoring. These tasks typically with small data sizes, ranging from kilobytes for sensing data to megabytes for short, low-resolution videos. The computation process of the monitoring task is intensive and encompasses various functions, including anomaly detection. Given the characteristics of monitoring tasks, they can be conducted by either ground edges or satellites with computation capability. The processed data will be then transmitted to a designated \textbf{ground station}, which is connected to cloud servers for further processing. The service \textbf{user} can collect the final processed results from the cloud servers if necessary. Based on the obtained results, the user may ask for confirmation of current monitoring information, i.e., a video streaming service is generated to transmit the high-resolution video data collected by a local camera to this user via STN. In contrast to monitoring tasks, video streaming tasks typically require more communication resources due to the extended duration of high-quality video streaming services.

%As illustrated in \figurename~\ref{fig:model}, the generated monitoring task that cannot be fully processed at the ground can also be assigned to the multiple satellites provided by the LEO constellation. The processed data will be then transferred to a designated station, i.e., local administration office for further processing. The processed data usually includes the information that whether the received task contents are normal. As a result, the size of data after processing is much more smaller than that of the original task. For simplicity, in this paper, we ignore the data communication for the final processed results since its impact on the overall network performance is limited. If there is no abundant resources left for data processing or the processing time exceeds the time limit of a computation task, this task will be aborted. \note{Based on the processing results, the users at the station may ask for confirming the current monitoring information, i.e., \textbf{a real-time video service} that transmits the video data collected by local camera to the user via STN.}

%and then transmits the processed data to the destination (i.e., cloud server) via LEO satellite constellation. %If there is no abundant resources left for data processing, the respective data will be transferred to the cloud server for further processing. The \textit{cloud server} stores the data processing results received from LEO satellites. The regional administrator nearby can then conduct further actions based on the logs of processing results.

\subsection{Communication model}~\label{subsec:cm}

\subsubsection{Edge to Gateway} In this subsection, we introduce the communication model for edge-to-gateway transmission. It is important to highlight that the communication model employed for edge-to-edge transmission is identical to the one used for edge-to-gateway transmission within a terrestrial network.

For each gateway, the edges within its coverage transmit the collected tasks to the respective gateway. The communication from edge to gateway is based on time-division multiple access (TDMA) to avoid interference during the data transmission. Assuming Gaussian channels, the uplink data rate from edge $e$ to gateway $g$ in the time $t$ is as below:%~\cite{ZhuJSAC23}:
\begin{equation}
    r_{g,e}^\textit{E2G} = B_{g}\log_{2}(1+\frac{{g_{g,e}^{t}}p_{g,e}}{B_{g}N_{0}}),
    \label{eq:dr_gnd}
\end{equation}
where $B_{g}$ is the channel bandwidth between gateway $g$ and edge $e$, $g_{g,e}^{t}$ is the channel gain from $g$ to $e$ in time $t$, $p_{g,e}$ is the transmission power, and $N_0$ is the noise power spectral density calculated by:
\begin{equation}
    N_0 = {k_B} T B,
\end{equation}
where $k_B$ is the Boltzmann constant, $T$ is the receiver noise temperature, and $B$ is the current channel bandwidth. It is worth noting that we assume gateways are isolated in remote or rural areas, hence there will be no interference between gateways. The maximum data rate $r^\textit{E2E}_{e,l}$ for edge-to-edge connection, e.g., edge $e$ to edge $l$, can be calculated in a similar way as Equation~\ref{eq:dr_gnd}.

\subsubsection{Gateway to Satellite}
Each gateway transmits the task to the satellite by TDMA, and similarly, the uplink data rate from gateway $g$ to the satellite in time  $t$ is:
\begin{equation}  
r_{g,i}^\textit{G2S} = B_s\log_{2}(1+\frac{{g_{e}^{t}}p_{e}}{B_{s} N_{0}}),
\end{equation} 
where $B_\textit{s}$ is the channel bandwidth, $g_{e}^{t}$ is the channel gain from gateway $g$ to the satellite in time  $t$, and $p_{e}$ is the transmission power.

\subsubsection{Satellite to Satellite}
Assuming the unobstructed line of sight by the Earth, the maximum data rate between satellite $k$ and $i$ is:
\begin{equation}
    r_{k,i}^\textit{S2S} = B_{s}\log_2(1 + \textit{SNR}(k,i)).
\end{equation}

Here, $\textit{SNR}(k,i)$ is the signal to noise ratio, which can be calculated by the following equation~\cite{LeyvaTWC21}:
\begin{equation}
    \textit{SNR}(k,i) =\frac{P_{t}{G_k(i)}{G_i(k)}}{{N_0}L(k,i)},
\end{equation}
where $P_t$ is the power of transmitter, $G_k(i)$ is the average antenna gain of satellite $k$ towards satellite $i$. $L(k,i)$ is the free space path loss calculated by:
\begin{equation}
    L(k,i) = \left(\frac{4{\pi}f_c{\textit{dist}(k,i)}}{c}\right)^2,
\end{equation}
where $f_c$ is the carrier frequency, $\textit{dist}(k,i)$ is the Euclidean distance between satellite $k$ and $i$, and $c$ is the speed of the light.

%The time for transmitting data $w$ from satellite $k$ to satellite $i$ can be calculated by:

\subsection{Computation model}

In the network, each device or user $d$ in the respective areas collects the sensing data and sends them to the designated edge $e$, where a set of computation tasks represented by $\mathbf{M}^g = \{m^g_1, m^g_2,...,m^g_n\}$ is generated. For each task $m\in \mathcal{M}$, it includes three information denoted as $\{D_m,C_m, T^\textit{max}_m\}$ , i.e., the data size $D_m$ for computation, the required computation resource (CPU cycle) $C_m$ and time limitation for task processing $T^\textit{max}_m$. If a task cannot be fully processed within the time limitation, this task will be discarded. We assume that each task is dividable, allowing us to employ the partial offloading mode~\cite{ZhuJSAC23}. 
Then, the edge can process the collected data as \textbf{local computing (LC)}. Let $\alpha_{m,e} \in [0,1]$ be the proportion of data processed at the edge, and thus the data size processed at edge $e$ is $\alpha_{m,e} D_{m,e}$ while the required computation resource is $\alpha_{m,e} D_{m,e} C_{m,e}$. 
% 这里的两个 \alpha_m_e  是一个意思吗，是一个意思的话，应该用一样的符号
Here, $D_{m,e}$ is the task input data size and $C_{m,e}$ is the required CPU cycles for computing the input data. If the current edge cannot fully process the computation task, i.e., $\alpha_{m,e} < 1$, this task will be transferred to a neighbor edge for processing. After conducting the local computing (if the whole processing time is within the time limitation for task processing), the remaining data size for a task $m$ is: 
\begin{equation}
    d_m^\textit{re,LC}= D_m - \sum_{\forall e}\alpha_{m,e} D_{m,e},
\end{equation}
and the remaining required computation resource $c_m^\textit{re,LC}$ is:
\begin{equation}
    c_m^\textit{re,LC} = C_m - \sum_{\forall e}\alpha_{m,e}D_{m,e}C_{m,e}.
\end{equation}

For each gateway $g$, when receiving a partially processed task $m$, i.e., both $c_{m}^\textit{re,LC}$ and $d_m^\textit{re,LC}$ is larger than 0, this gateway will offload the respective task to a selected LEO satellite for \textbf{satellite computing (SC)}. Assuming a proportion of data $\beta_{m,i} \in [0,1]$ can be processed at satellite $i$, and thus the data size processed at $i$ are $\beta_{m,i} D_{m,i}^\textit{SC}$ while the required computation resource is $\beta_{m,i} D_{m,i}^\textit{SC} C_{m,i}^\textit{SC}$. %Let $C_E^\textit{SAT}$ is the computation capacity of each satellite, $r_E^m$ is the computation resource allocated to the task $m$. In this case, $r_E^m \leq C_E^\textit{SAT}$. 
If task $m$ is fully processed at satellite $i$, it will not be offloaded to other available satellites. Otherwise, a portion of the computation task will be transferred to a neighbor satellite $k$ for further processing. After computing at one or several satellites, the remaining task size $d_m^\textit{re,SC}$ can be calculated by
\begin{equation}
    d_m^\textit{re,SC} = d_m^\textit{re,LC} - \sum_{\forall i}\beta_{m,i} D_{m,i}^\textit{SC}.
\end{equation}

If $d_m^\textit{re,SC} = 0$, the computation task is fully processed.

\subsection{Video streaming model}
A video is partitioned into a sequence of small segments, each containing a piece of the source video with a length $L_f$. These segments are then transferred to the designated UE on a segment-by-segment basis. In this paper, adaptive bitrate is utilized to provide high-quality video without introducing rebuffering capability on the UE side. This function is essential for the video streaming task considered in this work in terms of providing satisfactory QoE while maintaining high network performance. At each time $t$, a segment is encoded into multiple copies with various bitrates $r_t \in \mathcal{R}$, which allows the users to choose the most appropriate bitrate for each segment based on their QoE requirements and current channel conditions of STN. It is noteworthy that a higher bitrate level corresponds to superior video quality for each segment at the expense of utilizing additional network resources offered by STNs.

Based on \cite{WangJSAC22}, we define the energy consumption generated during the video streaming process as:
\begin{equation}
    e^p = e^e + e^u + e^c.  
    \label{eq:energy}
\end{equation}

Equation~\ref{eq:energy} includes three terms: $e^e$ is the encoding energy consumption denoted as:
\begin{equation}
    e^e = \kappa_v r^e L_f,
\end{equation}
where $r^e$ is the encoding bitrate and $\kappa_v$ is the energy efficiency parameter. $e^u$ is the transmission energy calculated by:
\begin{equation}
    e^u = \frac {t^u}{h}g(\frac{r_{t-1} L_f}{t^u})
    \label{eq:e_c}
\end{equation}
where
\begin{equation}
    g(x) = N_0W(2^{\frac{x}{W}}-1)
\end{equation}
In the above two equations, $t^u$ represents the uploading time, $W$ denotes the orthogonal channel bandwidth allocated to the streamer, and $N_0$ stands for the power spectral density of the noise. 

The third term $e^c$ is the energy consumption associated with video transcoding calculated by:
\begin{equation}
    e^c =\eta_v (f)^3 t^{tc},
\end{equation}
where $f$ is the CPU frequency, and $t^{tc}$ is the transcoding time.

\begin{comment}
and super-resolution are
incorporated into our design. \begin{itemize}
    \item \textit{Rate Adaptation}: Due to capacity constraints, the highest-resolution chunks cannot always be sent to the UE without rebuffering. To cope with this issue, the controller sends chunks at different bit rate levels according to the current channel conditions. The higher the bit rate level, the better the quality of the video chunk.

    \item \textit{Super Resolution}: To provide high-resolution chunks under fluctuating links, the controller can first send a low-resolution chunk to the relay satellite, upscale the chunk into a higher-resolution chunk on the satellite by a super-resolution algorithm, and then send the upscaled chunk to the UE. With super-resolution, we can trade transmission time for computation runtime, achieving a better QoE. The super-resolution runtime in our setting is different from [1] but consistent with [9]. The former assumes that the time complexity of super-resolution runtime increases with higher input resolution or higher output resolution, while the latter points out that higher output resolution will not always take longer super-resolution runtime, due to parallel computation and implementation. The following equation is used to calculate the super-resolution runtime of a chunk, $T_{sr,max}$.
    \begin{equation}
        T_{sr,max} = T_{sr,frame}\cdot T_c \cdot FR
    \end{equation}
    where $T_{sr,frame}$ is the super-resolution runtime of a frame and $FR$ is the video frame rate.

\end{itemize}
\end{comment}

%% file: 4_problem.tex
\section{Problem Formulation}~\label{sec:formulation}
\subsection{Delay analysis}~\label{subsec:delay_analysis}
\subsubsection{Local Computing}
For an arriving task $m$, three types of delay should be considered: Transmission delay $t^{LC}_\textit{tran}$, propagation delay $t^{LC}_\textit{prop}$ and computation delay $t^\textit{comp,LC}_{m,k}$ generated by computation process at edge $k$.

\begin{comment}
the delay in local computing executed at edge $e$ include the transmission delay $t^{LC}_\textit{tran}$, propagation delay $t^{LC}_\textit{prop}$ and computation delay at sender edge $k$ and computation delay at receiver edge $l$ as:
\begin{equation}
    t^\textit{sum,LC}_{m,k,l} = t^\textit{tran,LC}_{m,k,l} + t^\textit{prop,LC}_{m,k,l} + t^\textit{comp,LC}_{m,k}+t^\textit{comp,LC}_{m,l}
\end{equation}
\end{comment}

The transmission delay is calculated by dividing the size of data transmitted from edge $k$ to edge $l$ by the data rate between them:
\begin{equation}
    t^\textit{tran,LC}_{m,k,l} = \frac{\alpha_{m,k}D_{m,k}}{r^\textit{E2G}_{l,k}}.
\end{equation}
$r^\textit{E2G}_{l,k}$ herein can be calculated by Equation~\ref{eq:dr_gnd}. The propagation delay is the round trip delay by:
\begin{equation}
    t^\textit{prop,LC}_{m,k,l} = \frac{2\textit{dist}(l,k)}{c}
    \label{eq:propagation},
\end{equation}
where $c$ is the speed of the light. The computation delay at an edge $k$ can be determined by dividing the required computation resource by the computation capacity as:
\begin{equation}
    t^\textit{comp,LC}_{m,k} = \frac{\alpha_{m,k}D_{m,k}C_{m,k}}{f^\textit{EDGE}_{m,k}},
\end{equation}
where $f^\textit{EDGE}_{m,k}$ is the computation capacity of edge $k$ allocated to current task $m$. 

\begin{comment}
The total delay generated during local computing on task $m$ is:
\begin{equation}
    t^\textit{sum,LC}_{m} = \sum_{\forall k,l} t^\textit{sum,LC}_{m,k,l}
\end{equation}
\end{comment}

\subsubsection{Satellite Computing}
If a task $m$ is not fully processed by local computing and the current processing scale does not exceed the limitation, i.e., $d_m^\textit{re,LC} > 0$ and $t^\textit{tran,LC}_{m} < T^\textit{max}_m$, then additional delays must be considered. The delays include the transmission delay $t^{SC}_\textit{tran}$, propagation delay $t^{SC}_\textit{prop}$, and computation delay  $t^\textit{comp,LC}_{m,i}$ incurred by performing computation at satellite $i$. These delays can be calculated similarly to those for local computing. Additionally, the total delay encompasses the transmission and propagation delay for transferring the remaining computation task from edge $e$ to gateway $g$, as well as the transmission and propagation delay for transmitting the task from gateway $g$ to a reachable satellite $s$.
%If a task $m$ is not fully processed by local computing and the current processing scale does not exceed the limitation, i.e., $d_m^\textit{re,LC} > 0$ and $t^\textit{tran,LC}_{m} < T^\textit{max}_m$. In addition to the transmission delay $t^{SC}_\textit{tran}$, propagation delay $t^{SC}_\textit{prop}$, and computation delay  $t^\textit{comp,LC}_{m,i}$ incurred by performing computation at satellite $i$ (all of which can be calculated using a similar approach as local computing), the total delay also encompasses the transmission and propagation delay for transferring the remaining computation task from edge $e$ to gateway $g$, as well as the transmission and propagation delay for transmitting the task from gateway $g$ to a reachable satellite $s$.

\subsubsection{Total Delay}
The offloading decision to an arriving task $m$ depends on the task splitting ratio in local computing and $t^\textit{tran,LC}_{m}$. Assuming edge $k$ is the source node that generates the task. The total delay $t^\textit{total}_{m}$ is calculated by:
\begin{equation}
t^\textit{total}_{m}=\left\{
\begin{aligned}
&t^\textit{comp,LC}_{m,k}   ,  &\alpha_{m,k}=1, \\
&\max\{t^\textit{comp,LC}_{m,k}, t^\textit{comm,LC}_{m}\} ,  &\alpha_{m,k}<1, \\
&\max\{t^\textit{comp,LC}_{m,k}, t^\textit{comm,LC}_{m}+t^\textit{comp,SC}_{m,i},\\&\qquad\quad t^\textit{comm,LC}_{m}+t^\textit{comm,SC}_{m}\},   &\sum_{\forall e}\alpha_{m,e}<1,
\end{aligned}
\right.
\label{eq:total_delay}
\end{equation}

s.t.
    \begin{equation}
        t^\textit{total}_{m} \leq T^{\textit{max}}_m
    \label{eq:time_limit},
    \end{equation}
where 
\begin{subequations}
    \begin{equation}t^\textit{comm,LC}_{m} = \sum_{\forall k,l\in \mathcal{E}}(t^\textit{tran,LC}_{m,k,l} + t^\textit{prop,LC}_{m,k,l}),
    \label{eq:comm_lc}
    \end{equation}
    \begin{equation}t^\textit{comm,SC}_{m} = \sum_{\forall i,j\in \mathcal{N}}(t^\textit{tran,SC}_{m,i,j} + t^\textit{prop,SC}_{m,i,j})+t^\textit{comm,E2G}_{m} + t^\textit{comm,G2S}_{m}.\end{equation}
    \label{eq:comm_EC}
\end{subequations}
It is worth noting that the total delay cannot exceed the time limitation for processing the task, i.e., $T^{\textit{max}}_m$. If the delay value exceeds this limitation, the respective task will be discarded (Equation~\ref{eq:time_limit}). Considering this constraint, if $\alpha_{m,k}=1$, task $m$ will be fully processed at edge $k$ in terrestrial networks, and thus the total delay is the computation time on the edge. For the case where $\alpha_{m,k}<1$, indicating that the task needs to be offloaded to other neighbor edges, the total delay will be the greater of the computation time on an edge and the communication time among the edges (See Equation~\ref{eq:comm_lc}). Otherwise, when $\sum_{\forall e}\alpha_{m,e}<1$, task $m$ will be offloaded to satellites. Since local computing and satellite computing are processed simultaneously, the total delay can be calculated by the last part in Equation~\ref{eq:total_delay}. The communication time for satellite computing includes those required for edge-to-gateway connection, gateway-to-satellite, and satellite-to-satellite as shown in Equation~\ref{eq:comm_EC}.

\subsection{Resource usage analysis}~\label{subsec:ru_analysis}
\subsubsection{Local Computing}
If a computation task $m$ is processed at edge $k$, the computation resource usage is:
\begin{equation}
u^\textit{comp,LC}_{m,k} =\frac{f^\textit{EDGE}_{m,k}}{C^\textit{comp,LC}_k}, 
\end{equation}
where $C^\textit{comp,LC}_k$ is the computation capacity of edge $k$. Meanwhile, the communication resource usage is defined as: 
\begin{equation}
    u^\textit{comm,LC}_{m,k} =\frac{\alpha_{m,k}D_{m,k}}{r^\textit{E2E}_{k}}, 
\end{equation}
where $r^\textit{E2E}_{e}$ is the communication resource capacity (maximum data rate) obtained by using the equations introduced in Section~\ref{subsec:cm}. 

\subsubsection{Satellite computing}
Similar to the case for local computing, the computation resource usage for satellite $i$ that processes an arriving computing task $m$ is:
\begin{equation}
    u^\textit{comp,SC}_{m,i} =\frac{f^\textit{SAT}_{m,i}}{C^\textit{comp,SC}_i}, 
\end{equation}
where $f^\textit{SAT}_{m,i}$ is the computation capacity of satellite $i$ allocated to current task $m$ and $C^\textit{comp,SC}_i$ is the computation capacity of satellite $i$. Meanwhile, the communication resource capacity of satellite $i$ is:
\begin{equation}
    u^\textit{comm,SC}_{m,i} =\frac{\beta_{m,i}D^\textit{SC}_{m,i}}{r^\textit{S2S}_{i}}. 
\end{equation}

\begin{comment}
\subsubsection{Total Resource Usage}
The total resource usage for processing task $m$ includes those for computation resource and communication resource on both edges and satellites, which can be concluded as:
\begin{equation}
    u^\textit{total}_m = (u^\textit{comp,LC}_{m,k} + u^\textit{comm,LC}_{m,k,l})+ u^\textit{comm,G2S}_{m,g,s} + (u^\textit{comp,EC}_{m,i} + u^\textit{comm,EC}_{m,k,i} )
\end{equation}
\end{comment}

\subsection{Multi-objective optimization problem of task offloading}

%The task offloading includes the offloading decision to decide which device to execute the partitioned/whole task, and resource allocation to determine the amount of resources needed for processing the respective task. 
Task offloading involves two main steps: the offloading decision that determines the node(s) responsible for executing either the partitioned or entire task, and resource allocation, which identifies the amount of resources needed for processing the arriving task. As aforementioned, we assume that IoT devices $d$ collect the sensing data in remote or rural areas. The edge $e$ generates the computation task $m\in \mathcal{M}$ by aggregating the received data from $d$. If edge $e$ cannot fully process this task, the edge will send an offloading request to the gateway for \textbf{path selection}. Based on the characteristics of the arriving task and the resource usage of various elements in the terrestrial network and satellite network, the algorithm selects the edges and satellites for processing this task collaboratively. Specifically, when a gateway $g$ receives a request for a computation task $m$, it first searches the contactable edge and satellite candidates. From path candidate list $\mathcal{P}$ connecting the source node (edge) and destination node (edge or satellite), $g$ tries to find a path $p$ that minimizes the delay and the resource usage occupied by task $m$.

The main criteria to be optimized in this problem include the quality metric, delay, resource utilization, and energy consumption. The quality metric $q(r_m)$  is associated with the quality metric such as video multi-method assessment fusion (VMAF)~\cite{VMAF}, where $r$ indicates the selected video bitrate. As mentioned in Section~\ref{subsec:delay_analysis}, the delay includes those generated by task computation, data transmission, and data propagation, the value of which depends on the selected offloading decision. The value of delay $t^\textit{total}_m$ can be calculated by Equation~\ref{eq:total_delay}. Meanwhile, as described in Section~\ref{subsec:ru_analysis}, the resource usage is associated with edge and satellite, while the resource can be classified into two categories: computation resource and communication resource. The energy consumption $e_m^p$ is calculated by Equation~\ref{eq:energy} for each arriving task $m$. 

The routing and task offloading problem is formulated as: 

 \begin{equation}
 \begin{aligned}
        \max_{\mathcal{X}} \sum_{m \in \mathcal{M}}\sum_{p \in \mathcal{P}}
        \Bigl[ x_{m}^{p} &\Bigl( q(r_m)  - \alpha t^\textit{total}_m \\&-
        \gamma  u_m^\textit{comm} -  \beta u_m^\textit{comp}  - \delta e_m^p \Bigl )
        \Bigl],
    \label{eq:min}
\end{aligned}
\end{equation}

%  不改

\begin{comment}
 \begin{equation}
        \max_{\mathcal{X}} \sum_{m \in \mathcal{M}}\sum_{p \in \mathcal{P}}
        \left[ x_{m}^{p} \left( \note{q(r_m)}  - \alpha t^\textit{total}_m -\gamma  u_m^\textit{comm} -  \beta u_m^\textit{comp}  - \delta e_m^p  \right)
        \right],
    \label{eq:min}

\end{equation}

{\color{red}[PREVIOUS VERSION]
 \begin{equation}
        \min_{\mathcal{X}} \sum_{m \in \mathcal{M}}\sum_{p \in \mathcal{P}}
        \left[ x_{m}^{p} \left(  t^\textit{total}_m  +  \gamma  u_m^\textit{comm} +  (1 - \gamma ) u_m^\textit{comp} \right)
        \right],
    \label{eq:min}
    \end{equation}}
\end{comment}

s.t.
\begin{subequations}  
   \begin{equation}\sum_{p \in \mathcal{P}} x^p_m = 1, \quad \forall m \\\end{equation}
   \begin{equation}
        \gamma,u_m^\textit{comm},u_m^\textit{comp} \in [0,1], \quad 
 \forall m,
   \end{equation}

\end{subequations}
where
\begin{subequations}  
   \begin{equation}
      u_m^\textit{comm} = \max\{u^\textit{comm,LC}_{m,k}, u^\textit{comm,EC}_{m,i}\}, \quad \forall k,i\in S_{p},
      \label{eq:m_comm}
    \end{equation}
  \begin{equation}
    u_m^\textit{comp} = \max\{u^\textit{comp,LC}_{m,k}, u^\textit{comp,EC}_{m,i}\}, \quad \forall k,i\in S_{p}.
    \label{eq:m_comp}
  \end{equation}
\end{subequations}  
Regarding the constraints, $\mathcal{X}=\{x_m^{p}\mid m\in \mathcal{M}, p\in \mathcal{P}$\} denotes a binary variable to express whether a path $p$ is established for task $m$, i.e., if $p$ is selected for accommodating task $m$, $x_m^{p}=1$, otherwise, this value is 0. For each $p$, it is essential to find the satellites for data processing. We denote $S_p$ as all nodes in the network traversed by $p$. Equations~\ref{eq:m_comm} and~\ref{eq:m_comp} indicate the maximum utilization of communication resources and computation resources traversed by path $p$ assigned to task $m$, respectively. 

Because the offloading decisions associated with the coupling relationship among different variables, even the single-objective optimization problem is a discrete and
non-convex optimization problem, which is a classical NP-hard problem~\cite{TangIoT21,ChenJSAC23}.

%% file: 5_algorithm.tex
\section{Collaborative Computing Algorithm for Adaptive Video Streaming in STNs}\label{sec:algorithm} 
\subsection{Problem decomposition}~\label{subsec:framework}

\begin{figure*}[tb!]
    \centering
    \includegraphics[width=1\linewidth]{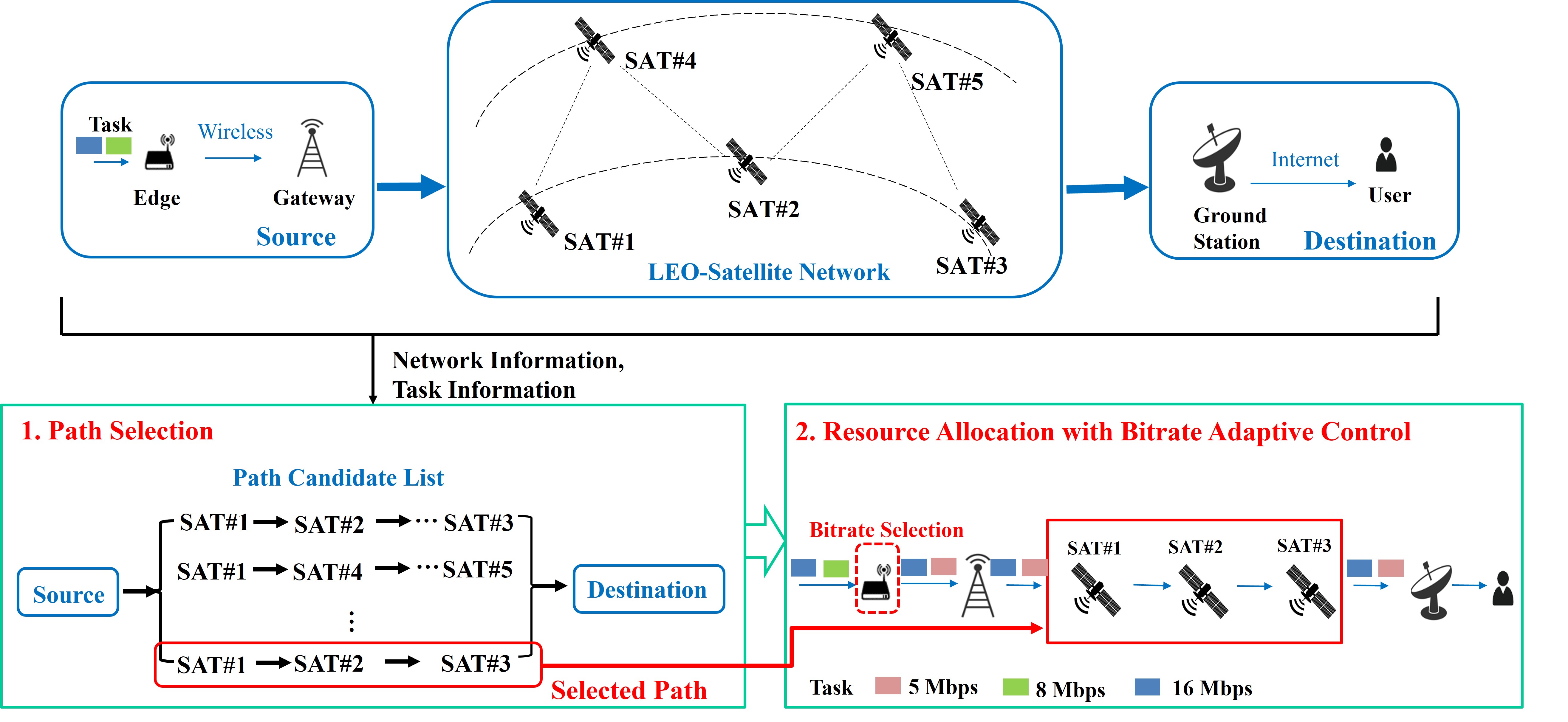}
    \caption{Overview of the proposed algorithm. }
    \label{fig:alg}
\end{figure*}

To solve this problem, we develop efficient algorithms with much lower computational complexity for joint task offloading. We decompose the original problem into two sub-problems: path selection sub-problem and resource allocation with adaptive bitrate control sub-problem as illustrated in \figurename~\ref{fig:alg}. 

For the path selection sub-problem, it dynamically selects an end-to-end path from multiple potential candidates based on the current network condition and task information. The objective of this sub-problem is to minimize the service delay while maintaining a high resource utilization in terms of computation and communication resources. The objective is formulated as below:
\begin{equation}
        \min_{\mathcal{X}} \sum_{m \in \mathcal{M}}\sum_{p \in \mathcal{P}}
        \Bigl[ x_{m}^{p} \left( t^\textit{total}_m + \gamma  u_m^\textit{comm} +  \beta u_m^\textit{comp}    \right)
        \Bigl].
    \label{eq:min1}
    \end{equation}

The resource allocation with adaptive bitrate control sub-problem focuses on bitrate adaption optimization over the path selected in the previous sub-problem. %In this sub-problem, both network information and task information (See \figurename~\ref{fig:alg}) are used to assist the decision making process. 
The objective of this sub-problem is to maximize QoE and task completion ratio $r^c$, while reducing the energy consumption $e_p$ as:
\begin{equation}
        \max \Bigl[\sum_{m \in \mathcal{M}}
        \Bigl( q(r_m)  - \delta e_m^p \Bigl) + \omega r^c\Bigl],
    \label{eq:obj2}
\end{equation}
where QoE is defined as~\cite{ZhangINFOCOM19}: 
\begin{comment}
\begin{equation}
q(r_m) = \sum_{\forall m \in \mathcal{M}_{t+1}}q(r_{m,t+1}),
\label{eq:QoE}
\end{equation}
where
\end{comment}
\begin{equation}
    q(r_m) = \eta r_{m,t+1} - \kappa |r_{m,t+1}-r_{m,t}| - \nu \frac {r_{m,t+1}}{f_{t+1}}.
    \label{eq:QoE}
\end{equation}

The parameters $\eta$, $\kappa$, and $\nu$ represent user preferences. The variable $r_{m, t+1} \in \mathcal{R}$ denotes the selected bitrate for task $m$ and $f_{t+1}$ is the link usage at time $t+1$. The three terms in Equation~\ref{eq:QoE} show the video playback quality, the quality degradation loss, and the time lost due to re-buffering, respectively. The task completion ratio $r^c$ is defined as the ratio of tasks successfully processed at time $t$, while the energy consumption $e_m^p$ is obtained by Equation~\ref{eq:energy}.

\subsection{Path selection}~\label{subsec:path_selection}

%\subsection{Information sharing}
As a preparation stage, each satellite needs to obtain the following information from ground stations in advance or from the neighbor satellites on a hop-by-hop basis:

\begin{itemize}
    \item \textit{Topology graph information} $T_s$ indicates the relative location of each satellite to a given satellite $s$. By using this information, the path between $s$ and any other satellites can be calculated. % obtained from neighbors on a hop-by-hop basis. The neighbors can be adjacent satellites located in a given communication range of satellite $s$. In a given interval, each satellite $s$ broadcast link state as $\textit{LS}_s$ = \{neighbors: $\emptyset$, hop count: 0\} to its neighbors $s\prime$, then $s\prime$ transfers the updated information as $\textit{LS}_s$ = \{neighbors: $s\prime$, hop count: 1\}, \note{this broadcast is executed until each satellite} 
    \item  \textit{Satellite-user pair information} \textit{$P$} tells which satellite is connectable to the designated user, by which the destination satellite of a path can be specified. 
    %\item {\color{blue}[OPTIONAL?]}\textit{CL}: Congestion link information includes the LEO satellite links which needs to be excluded for newly establishing paths to avoid further link congestion. When a satellite detects its communication link is congested, it broadcasts the congestion information to other LEO satellites. The definition of congestion can be decided by a given threshold or the current congestion status compared with the link usage of other neighbors.%, \note{or bio-inspired method (See Appendix).}
\end{itemize}

\begin{figure}[!t]
\removelatexerror
\begin{algorithm}[H]
\SetKwInput{Input}{Input}
\SetKwFor{For}{for}{}{end}
\SetKwFor{When}{when}{}{end}
\SetKwFor{Foreach}{for each}{}{end}
\SetKw{GoTo}{go to}
\SetKw{Return}{return}
\caption{Path Selection Scheme}\label{Alg:joint}
\Input{ Satellite $s$ on which the algorithm is running\; }%, $C_l$: link resource capacity, $C_e$: computational resource capacity\; }

\When{task $m$ arrives to send DATA to user \textit{u}}
{
    Search $(i, \textit{ls})$ from $P$\; \label{line:gs}
		%Initialize $u_s\leftarrow 0$  \;
	
	\For{$\textit{count}$ $\gets 1$ to $\textit{COUNT}_\textit{max}$ } 
	{
	    
	   Compute a new shortest path $p$ to $i$ by $T_{s}$ \;  \label{line:rc}
	   \If{ Path reservation for $p$ is accepted \label{line:check}} 
	   {    
	        %Collect the following information from 	$p$:\newline 
	        %$k \leftarrow$ satellite ID (= hop distance $hop(k)$ to $s$) \newline
			%$u_E^k \leftarrow$ available computational resource usage of satellite $k$, \newline 
	        %$u_L^k \leftarrow$ available communication resource %usage of link between satellite $k$ and $k+1$ \label{line:check1} \;
	        Compute candidate score by \textbf{\Algtwo}\; \label{line:alg}

			%\If{$u_s < u_p$ }
			%{
			%    $u_s \leftarrow u_p$, $p_s \leftarrow p$, $k_s \leftarrow k$\;
			%}
	    }
	     \Else{
	    	Remove the congested link from $T_{s}$\;
	    }

	}
	\If{at least one path candidate is found}
	{
	Select path $p_s$ with the best score\;
	Send \textit{DATA} to user \textit{u} via $p_s$\;
	Assign designated satellites for data processing\;
	}

}

 \end{algorithm}
\end{figure}

Algorithm~\ref{Alg:joint} summarizes the flow of path selection scheme. When a satellite $s$ receives a data connection request from the gateway, it first confirms a destination satellite $i$ that is connectable to the designated user by searching $P$ (line~\ref{line:gs} in Algorithm~\ref{Alg:joint}), and then computes the shortest path candidate from $s$ to $i$ based on $T_s$. For each path candidate, $s$ tries to reserve link resources for data connection, i.e., by using resource reservation protocol (RSVP) or other associated mechanisms~\cite{LanWCNC21,QiuHotICN22}, while obtaining the communication and computation resource usage of the current path candidate.

As depicted in line~\ref{line:alg} in Algorithm~\ref{Alg:joint}, we introduce a heuristic approach named path selection based on the resource usages (\textbf{\Algtwo}). This algorithm scores each path candidate to select the ideal one for data connection. It aims to assign the task to a path candidate with the most abundant available resources while minimizing propagation delay. For each path candidate $p$, it calculates the cost value $s^p_\textrm{psru}$ using the following equations:

\begin{equation}
\label{eq:rsc}	
		s^p_\textrm{psru} =\alpha {\bar{r}_L^p} + (1-\alpha) {\bar{r}_E^p}.
\end{equation}
where 
\begin{subequations}  
\begin{equation}
\label{eq:rsc_link}
    		\bar{r}_L^p =   \frac{1}{\textit{len}(p)}\sum_{\forall k }\left(\frac{ u_L^k}{C_L}\right),
\end{equation}	
\begin{equation}
\label{eq:rsc_edge}
    	\bar{r}_E^p =  \frac{1}{\textit{len}(p)}\frac{u_E^k}{C_E}.
\end{equation}
\end{subequations}

%We set $\alpha$ = 0.5 which is the weight on the computation resource  (See Equation~\ref{eq:rsc}). 
Equations~\ref{eq:rsc_link} and \ref{eq:rsc_edge} indicate the current vacant resources of data commutation and computation respectively. The variable $u^k_L$ denotes the available communication resource for the link connecting satellite $k$ to its next hop, while $u^k_E$ is the available computation resource on satellite $k$. The path with a maximum $s^p_\textrm{psru}$ value is then selected for data connection. %For all tasks, this algorithm needs to search all path candidates with a number $\textit{COUNT}_\textit{max}$ and this process requires $\frac{\textit{COUNT}_\textit{max}N(N+1)}{2}$ comparisons where $N$ is the total number of data connection tasks. Therefore, assuming $N$ as the total number of data connection tasks, the computational complexity is $O(N^2)$ which is affordable for dynamic ground-space integrated networks~\cite{ChengJSAC19}.

%\subsection{Resource allocation}

\subsection{Resource allocation with adaptive bitrate streaming control}~\label{subsec:masac}

\begin{figure}[tb!]
    \centering
    \includegraphics[width=1\linewidth,trim=120 35 50 40,clip]{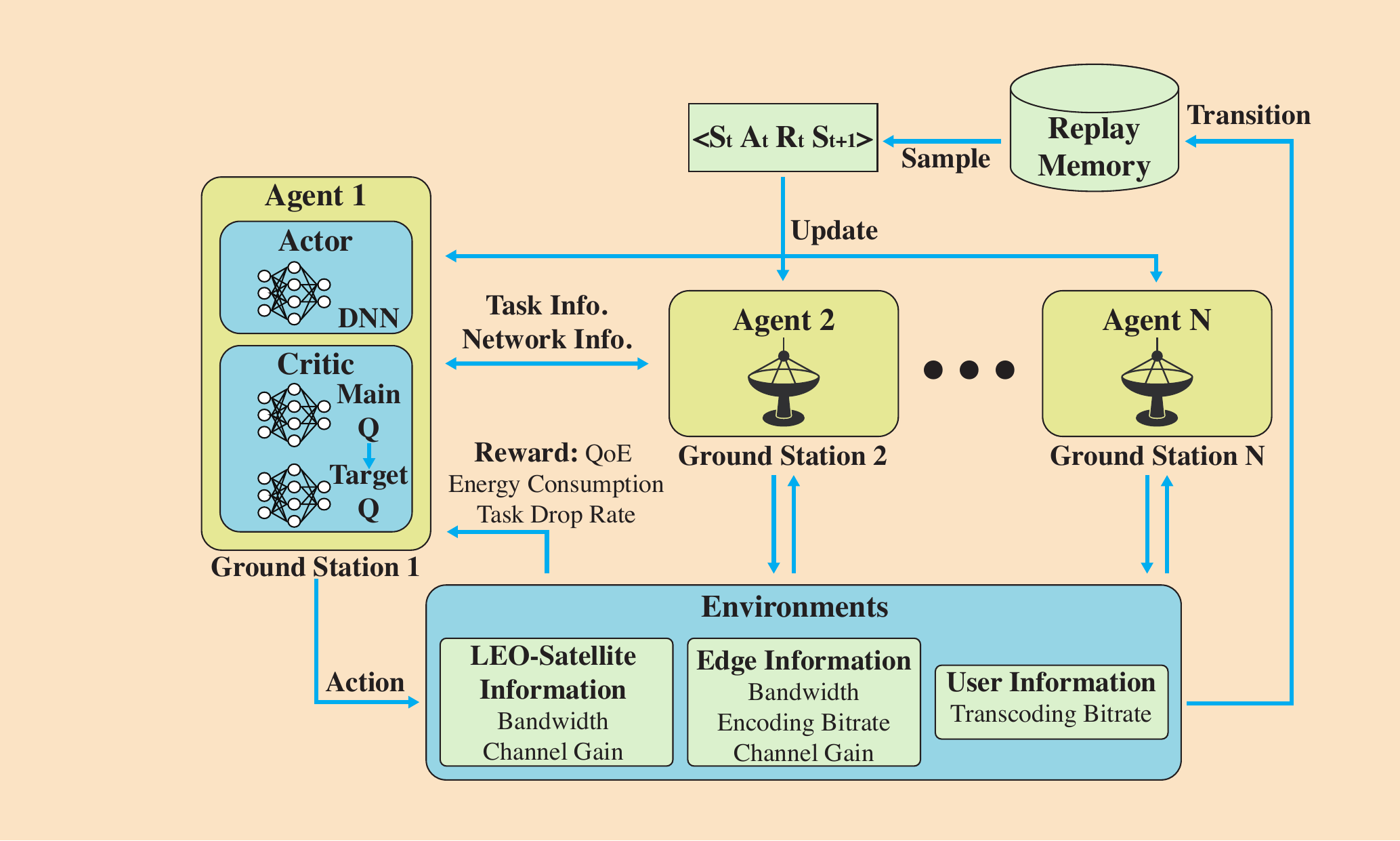}
    \caption{Resource allocation with adaptive bitrate streaming control based on MA-SAC. }
    \label{fig:sac}
\end{figure}

We design a resource allocation algorithm with online adaptive video streaming control to maximize service QoE and task processing rate while reducing energy consumption. As illustrated in \figurename~\ref{fig:sac}, this algorithm determines the bitrate for incoming video streaming tasks based on the resource usage of multiple edges and satellites traversed by the path candidate selected in the previous stage. We assume that the initial edge receiving the video streaming task as the bitrate decision point. This is because all computing tasks must undergo preprocessing, including video compression, at this node to optimize the format, thereby reducing the network's bandwidth requirements. An agent is defined as the ground station connecting between satellite and users. Each independent agent to make its own decision autonomously, while achieving the common objective (Equation~\ref{eq:obj2}). We transform the sub-problem to a Markov decision process (MDP) as below:

\subsubsection{State space}
The environment state $s_t$ consists of the number of agents, the information of all agents including the encoding bitrate, transcoding bitrate, and the bandwidth occupied by the task $m$. 

\subsubsection{Observation space} At each time step $t$, each agent $i$ receives observed information associated with satellites, edges and users (See \figurename~\ref{fig:sac}). In details, the information from both the satellite and edge that includes bandwidth and channel gain status, with the edge node also providing encoding bitrate details. Additionally, user information encompasses the transcoding bitrate.

%bandwidth and channel gain status are included in the information of both satellite and edge, while encoding bitrate further includes in edge information. Meanwhile, user information includes the transcoding bitrate. 

%$o_{i,t}$ = $\{I_{m,t}, \mathcal{U}^\textit{comp}_{p,t}, \mathcal{U}^\textit{comm}_{p,t}\}$, 
%where $I_{m,t}$ denotes information related to task $m$.   $\mathcal{U}^\textit{comp}_{p,t}$ and $\mathcal{U}^\textit{comm}_{p,t}$ represent the set of the currently available computational and communication resource utilization of all nodes that path $p$ traverses, as determined by the previous path selection process.
\begin{comment}
\begin{equation}
    O_m = \{I_m, \mathcal{U}^\textit{comp}_p, \mathcal{U}^\textit{comm}_p \}
\end{equation}
\end{comment}

\subsubsection{Action space}
The action is defined as $a_t = \{r_{m,t+1}\}$ to determine bitrate for the arriving task $m$. Based on the assigned bitrate, then it is necessary to allocate the computation and communication resource under the current resource capacity limitation.

%The action space is defined as $\{\mathcal{r}^\textit{comp}_{m,p},\mathcal{r}^\textit{comm}_{m,p} \}$,
\begin{comment}
\begin{equation}
    A_m = \{\mathcal{r}^\textit{comp}_{m,p},\mathcal{r}^\textit{comm}_{m,p} \} = \{{r}^\textit{comp}_{m,p,1},...{r}^\textit{comp}_{m,p,I},{r}^\textit{comm}_{m,p,1} ..., {r}^\textit{comm}_{m,p,I}\}
\end{equation}

\begin{subequations}
\begin{equation}
\sum{{r}^\textit{comp}_{m,p,i}}=1,\sum{{r}^\textit{comm}_{m,p,i}}=1
\end{equation}
\begin{equation}
   {r}^\textit{comp}_{m,p,i},{r}^\textit{comm}_{m,p,i}\in [0,1]
\end{equation}
\end{subequations}
\end{comment}
%where $\mathcal{r}^\textit{comp}_{m,p}$ and $\mathcal{r}^\textit{comm}_{m,p}$ represent the set of  allocated computational and communication resource ratio (the value ranges between 0 and 1) on each node $i$ traversed by $p$.

\subsubsection{Transition probability}
$\mathcal{P}$: $p(s_{t+1}|s_t,a_t)$ is the probability of transition from state $s_t$ to $s_{t+1}$ when all the agents take joint action $a_t$=\{$a_{i,t}\}_{i=1}^N \in \mathcal{A}$.

\subsubsection{Reward function}
Agent $i$ obtains a reward $r_i^t$ by reward function $\mathcal{S}\times\mathcal{A}_1\times \cdots \times\mathcal{A}_N\rightarrow \mathcal{R}$. After conducting action $a_m$ under state $s_m$ at each time, the environment will enter a new updated state and return its corresponding reward. The reward function is formulated based on the optimization objective defined by Equation~\ref{eq:obj2}.

\subsubsection{Policy} 
At time $t$, agent $i$ chooses an action $a_t$ based on a certain policy $\pi_t$, aiming to maximize the total reward.  

We introduce an algorithm based on MA-SAC to obtain the optimal policies by learning the corresponding action functions through interaction with the environments. In this algorithm, the agent learns to conduct collaborative resource allocation among various components in STNs. Each agent considers the actions of various agents during the policy update process. The optimal policy is determined according to the trained model by MA-SAC. The objective is to optimize the cumulative expected rewards while maximizing the expected entropy of the policy as:
\begin{equation}
    \pi^{*} = \arg\max_{\pi}\mathbb{E}_{\pi}\left(\sum_{\forall t}r(s_t,a_t)+\alpha_H H(\pi_t(\cdot|s_t))\right),
\end{equation}
where $\alpha_H$ is a trade-off coefficient and $H(\pi_t(\cdot|s_t))$ is the entropy calculated by:
\begin{equation}
    H(\pi_t(\cdot|s_t))=-\sum_{a}\pi_t(a_t|s_t)\log \left(\pi_t(a_t|s_t)\right).
\end{equation}

Algorithm~\ref{alg:SAC} shows the details of our developed MA-SAC algorithm. It consists of two phases: experience gathering and network training. In the experience gathering phase, the experience replay mechanism is incorporated to mitigate the correlation among data samples. Each agent $i$ performs the actions $a$ generated in each episode, and then stores the tuples $(s_t,a_t,r_t,s_{t+1})$ into the replay buffer (line \ref{line:select} to \ref{line:store} in Algorithm~\ref{alg:SAC} and \figurename~\ref{fig:alg}).

\begin{algorithm}[ht!]
\SetKwInput{Input}{Input}
\SetKwInput{Output}{Output}
\SetKwFor{For}{for}{}{end}
\SetKwFor{When}{when}{}{end}
\SetKwFor{Foreach}{for each}{}{end}
\SetKw{GoTo}{go to}
\SetKw{Return}{return}
\SetKw{Repeat}{Repeat}
\SetKw{Until}{Until}{}
\caption{Multi-agent Soft Actor-Critic (MA-SAC)}\label{alg:SAC}
%\Input{ \;}
%\Output{\;}

Initial policy parameters $\theta$, Q-function parameters $\phi_{1}$, $\phi_{2}$, empty replay buffer $\mathcal{D}$\;
Set target parameters equal to main parameters $\phi_{\textit{targ}.1}\leftarrow \phi_{1}$, $\phi_{\textit{targ}.2}\leftarrow \phi_{2}$\;
\Repeat
Observe state $s$ and select action $a \sim \pi_\theta(\cdot |s_t)$\;\label{line:select}
Store ($s_t,a_t,r_t,s_{t+1}$) in replay buffer \textbf{$\mathcal D$}\;\label{line:store}
%\note{If s' is terminal, reset environment state\;}
%\If{\note{$D_\textit{size} > \textit{Thres}_\textit{size}$}}{
   \For{each training step }{
       Sample a random mini-batch of transitions, $\mathcal{B}= \{(s_t,a_t,r_t,s_{t+1}) \}$ from \textbf{$\mathcal D$}\;
       Compute targets for the Q functions by Equation~\ref{eq:target}\;
       Update critic networks by Equation~\ref{eq:loss}\;%one step of gradient decent using:
       %$\nabla_{\phi_i}\frac{1}{|B|}\sum_{(s,a,r,s') \in \mathcal{B}}{(Q_{\phi_i}(s,a)-y(r,s',d))^2}$\  for $i$=1,2\;
       %Update policy by one step of gradient ascent using Equation~\ref{eq:policy_update}\;
       Update target networks with $\phi_{targ,1}\leftarrow \rho\theta_{\textit{targ},1} + (1-\rho)\phi_1$,  
       $\phi_{targ,2}\leftarrow \rho\theta_{\textit{targ},2} + (1-\rho)\phi_2$~\label{line:targetnw}\;
   }
  %}
\Until{convergence}

\end{algorithm}

During each training step, a subset of data is randomly sampled from the replay buffer in order to update the parameters of both the actor networks and critic networks. 
\begin{comment}
The actor network updates the target by:
\begin{equation}
\begin{split}
    \mathcal{J}  =&  \mathbb{E}(\alpha \log(\pi_i(a_i|s_i)) \\&-Q_i^{\pi}(s_t,a_1,...,a_n)|_{a_i=\pi_i(s_i)}), i=1,2.
    \end{split}
\end{equation}
\end{comment}
The target for Q functions is expressed by:
\begin{equation}
      \begin{split}
       y_i &=  r_i + \\&\gamma \mathbb{E}\left(\min_{i=1,2}{Q_{\phi_{\text{targ},i}}{(s_{t+1},\tilde{a}_{t+1})}} - \alpha \log \pi_{\theta}(\tilde{a}_{t+1}| s_{t+1})\right),\\& \tilde{a}_{t+1} \sim \pi_{\theta}(\cdot| s_{t+1}).
       \label{eq:target}
       \end{split}
       \end{equation}

Based on the obtained targets, the critic networks are updated by minimizing the loss function calculated by:
\begin{equation}
    \mathcal{L} =
    \mathbb{E}_{s_t,a_t,r_t,s_{t+1}}\left(Q_i^{\pi}(s_t,a_1,...,a_n)-y_i\right)^2.
    \label{eq:loss}
\end{equation}

To ensure the training stability, the parameters of both actor networks and critic networks are copied to the corresponding target networks by soft update method as shown in line~\ref{line:targetnw} of Algorithm~\ref{alg:SAC}.

%SAC concurrently learns a policy \pi_{\theta} and two Q-functions Q_{\phi_1}, Q_{\phi_2}. There are two variants of SAC that are currently standard: one that uses a fixed entropy regularization coefficient \alpha, and another that enforces an entropy constraint by varying \alpha over the course of training. For simplicity, Spinning Up makes use of the version with a fixed entropy regularization coefficient, but the entropy-constrained variant is generally preferred by practitioners.

\begin{comment}
\cpy{ Update policy by one step of gradient ascent using
\begin{equation}
     \nabla_{\phi_i}\frac{1}{|B|}\sum_{s \in \mathcal{B}}{min_{i+1,2}(Q_{\phi_i}(s,\tilde{a}_{\theta}(s))-\alpha\log\pi_{\theta}(\tilde{a}_{\theta}(s)|s)},
       \label{eq:policy_update}
\end{equation}
where $\tilde{a}_{\theta}(s)$ is a sample from $\pi_{\theta}(\cdot|s)$, which is differentiable wrt $\theta$ via the reparametrization trick.}
\end{comment}

%\subsection{Complexity Analysis\note{??}}

%% file: 6_experiment.tex
\section{Experimental Evaluation}~\label{sec:evaluation}
\subsection{Experimental setup}

\begin{figure}[tb!]
    \centering
    \includegraphics[width=\columnwidth,trim=100 130 100 200,clip]{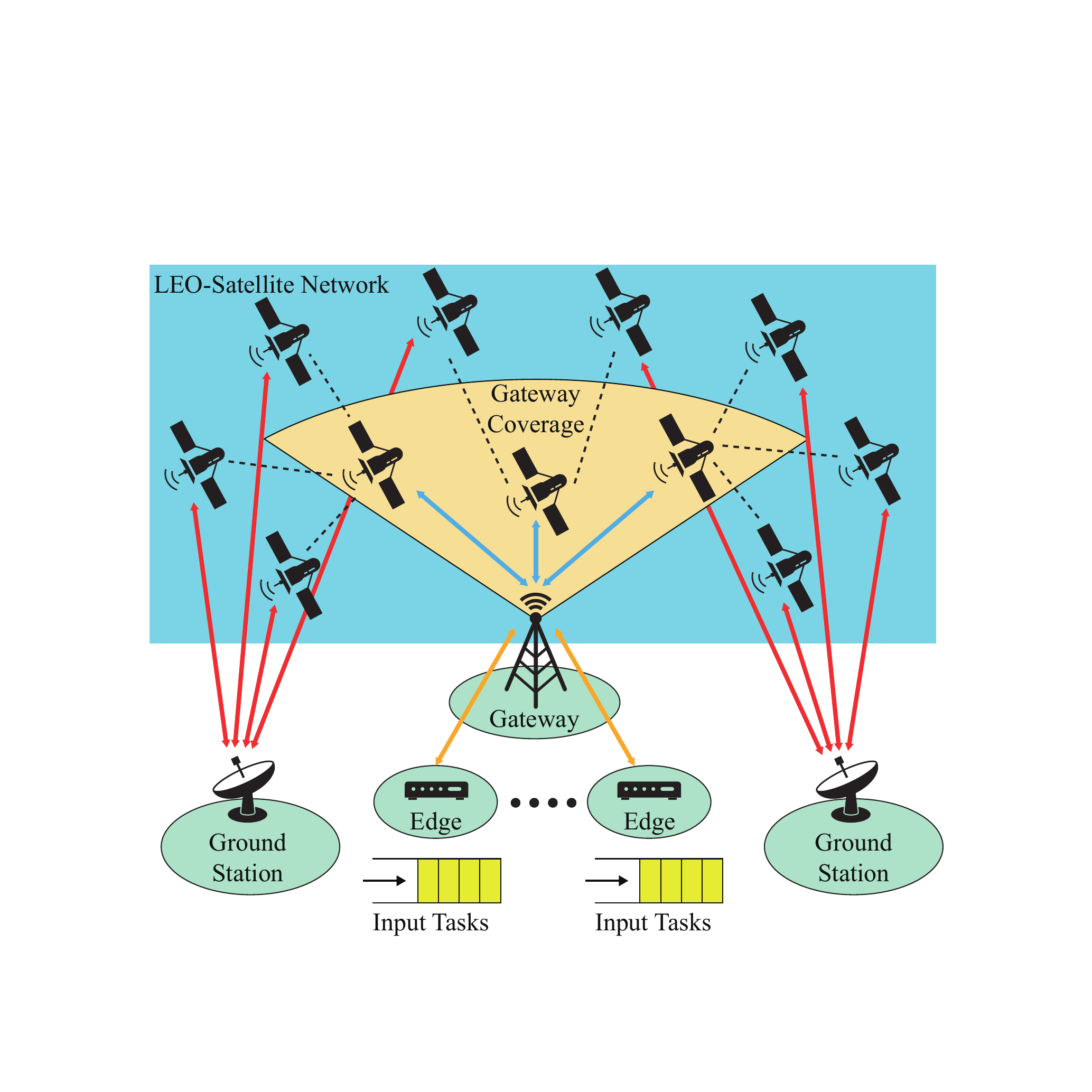}
    \caption{Experimental network topology.}
    \label{fig:setup}
\end{figure}

We build a simulation platform for an STN as shown in \figurename~\ref{fig:setup} for performance evaluation. The computing tasks are generated from 10 edges, each managed by a gateway. The gateway is connected to a LEO satellite network composed of three interconnected core satellites (indicated by the blue lines in \figurename~\ref{fig:setup}). Meanwhile, each core satellite has the capability to connect with two or three neighbor satellites for task offloading (shown by the dash lines in \figurename~\ref{fig:setup}). The satellite network is also linked to 3 ground stations (the red lines in \figurename~\ref{fig:setup}), which in turn connect to the local users.

As mentioned in Section~\ref{subsec:sysmodel}, we assume two types of computing tasks: monitoring and video streaming. The data size for monitoring tasks is randomly selected from a range of 0.1 MB to 0.5 MB. For video streaming tasks, the data size is drawn from the set [1.28, 3.2, 5.12, 7.68] MB, corresponding to four available bit rate levels: 16 Mbps (2K), 8 Mbps (1080P), 5 Mbps (720P), and 1 Mbps (360P)\cite{ZhangINFOCOM19}. The number of monitoring tasks and video streaming tasks are generated equally. Other main parameters used in the experiment are summarized in Table~\ref{tab:params}. 

\begin{comment}
\begin{figure}[tb!]
    \centering
  \subfloat[Satellite global distribution]{
      \includegraphics[width=0.7\linewidth]{Figure/P1.jpg}}
    \hfill\newline
  \subfloat[Satellite distribution around Victoria region (southeast part of Australia)]{
       \includegraphics[width=0.7\linewidth]{Figure/P2.jpg}}
  \caption{Captured information of Starlink satellite's position. \note{[NOT REQUIRED]}}
  \label{fig:sat_info} 
\end{figure} 
\end{comment}

\begin{table}[tb]
 \caption{Main experimental parameters}
 \label{tab:params}
 \centering
\begin{tabular}{ll}
   \hline
   Parameter  & Value \\
   \hline \hline
   %Satellite number $|\mathcal{N}|$ & \note{12??}\\
   %Orbit number & 20 \note{(different variation)}\\
   %Satellite number per orbit & 20 \note{(different variation)}\\
   Satellite bandwidth $B_s$& 27~GHz (Ka-band)\\
   %Satellite datarate $r^\textit{S2S}_{k,i},r^\textit{E2S}_{k,g}$ & \note{?}\\
   %Satellite link capacity & \note{120~Mbps (L-band)}\\
   %\note{ACHTUNG: Upgrade auf Ka-band} & \\
   %Altitude\note{?} &  1,150~km \\
   %Indication\note{?} & 53$^{\circ}$ \\
   Satellite computation capability $f^\textit{SAT}_{m,i}$& 50 GC/s (GC = $10^9$ cycles)~\cite{ZhuJSAC23}\\
   Satellite transmission power $P_t$
   & 30~dBW~\cite{ZhangIoT23}\\
   Noise power density $N_0$& -174~dBm/Hz~\cite{ZhuJSAC23,ChenJSAC23}\\
   Satellite antenna gain $G_k(i)$ & 16.2~dBi~\cite{LinCM21}\\
   %Edge number $|\mathcal{E}|$  & \note{???}\\
   Edge computation capability $f^\textit{EDGE}_{m,k}$& 0.5 GC/s~\cite{ZhuJSAC23}\\
   Ground node bandwidth $B_g$ & 4~GHz (C-band)\\

   %Ground node datarate $r^\textit{E2G}_{g,e},r^\textit{E2E}_{g,e}$\\
   
   Parameters related to MA-SAC & Agent\_num=3, others follow ~\cite{WangJSAC22}\\
   
   Task computation data ratio:\\
   --Monitoring task &  200 cycles/byte \\

   --Video streaming task&  50-100 cycles/byte \\
   
   %Parameters related to path selection & $\textit{COUNT}_\textit{max}$=\note{5?}\\

   \hline
  \end{tabular}
\end{table}

We  verify  the  effectiveness of  our proposed collaborative computing MA-SAC scheme (\textbf{\algname}) with the following four comparative schemes:\begin{itemize}

    %\item \textbf{{Local Computing (LC)}}: All arriving computation task can only be processed at the edges at the ground.

    %\item \textbf{\note{LC-MAIN??}}~\cite{ZhangIoT23}: For this approach, a small-scale satellite network ($|\textrm{N}|$ = 5) is used to assist the computation task offloading, which is calculated by DDPG. 

    \item \textbf{{Random-max-bitrate (RND-MAXBR)}}: It randomly selects a path with the available resources for task offloading~\cite{HanIoTJ23}.
    % It selects the maximum available bitrate for video streaming tasks. 
    % 代码中只是路径是随机选择，其他的没有改动，所以就把这句话注释掉了

    %\item \textbf{{First-fit (FF)}}: It selects the first found shortest path with the available resources for task offloading.

    \item  \textbf{Residual-resource-priority (RRP)}: Compared with the previous RND-MAXBR, this scheme selects the available edges and satellites with the most residual communication and computing resources for task offloading. %It also uses the maximum available bitrate for video streaming tasks. 

   %\item \textbf{Proximal policy optimization (PPO)} introduces PPO to obtain the optimal decision in adaptive bitrate streaming with resource allocation.

    %\item  \textbf{SAC} \note{[OPTIONAL]}: It a SAC-based algorithm modified from~\cite{DuSIGCOMM20}, which aims to obtain the optimal decision that maximizes the defined reward for adaptive bitrate streaming with resource allocation.

    \item \textbf{Deep deterministic policy gradient (DDPG)}: It is a DDPG-based computing offloading strategy~\cite{ZhangIoT23}. DDPG adds the target network and soft update methods, which plays a key role in the stable learning of the value network and policy network constructed by the deep learning model.
    
    \item \textbf{SACCT}: It is a SAC-based framework integrated with communication transformers~\cite{WangJSAC22}, which considers the uplink bitrate adaptation and edge transcoding for adaptive bitrate streaming.

    %\item \textbf{{DQN}}~\cite{mnih2015humanlevel}: As the mostly commonly used DRL algorithm for computing offloading~\cite{ZabihiCSUR23}, DQN is applied for resource allocation. In DQN, $\epsilon$-greedy policy is used for exploration during the training process. The network architecture of DQN is the same as the critic network of DDPG used in \algname. 

    %\item \note{DDQN [???]}
    %\item \note{PPO}

    %\item \note{TD3}

\end{itemize}

%\note{Comparison methods: Similar as ICC22 + ABR-related, PPO?, SAC,}

\subsection{Experimental results}

\begin{comment}
\begin{figure*}[t]
	\centering
	\begin{minipage}[b]{.666\columnwidth}
		\centering
		\includegraphics[width=\columnwidth]{Figure/convergence.jpg}
		\subcaption{Reward}\label{fig:t_reward}
	\end{minipage}
        \begin{minipage}[b]{.666\columnwidth}
		\centering
		
		\includegraphics[width=\columnwidth]{Figure/qoe.png}
		\subcaption{QoE}\label{fig:t_qoe}
	\end{minipage}
	\begin{minipage}[b]{.666\columnwidth}
		\centering
		\includegraphics[width=\columnwidth]{Figure/convergence.jpg}
		\subcaption{Energy consumption}\label{fig:t_energy}
	\end{minipage}
		
	\caption{Convergence performance.\note{To be UPDATED}.}
	\label{fig:convergence}
\end{figure*}
\end{comment}

First, we evaluate the training performance for different SAC-based schemes. Since the original SACCT was initially designed for 2-hop MEC networks, rather than the STNs assumed in this paper, the same path selection scheme to our proposal has been implemented for SACCT to ensure a fair comparison. We set the number of tasks to be trained in each episode to 25 with the batch size 256 for all schemes during the training process. Other main parameter settings for the DRL training follow \cite{WangJSAC22}.

\begin{figure}[t]
\centering
\subfloat[Reward]
{\includegraphics[scale=0.48,clip]{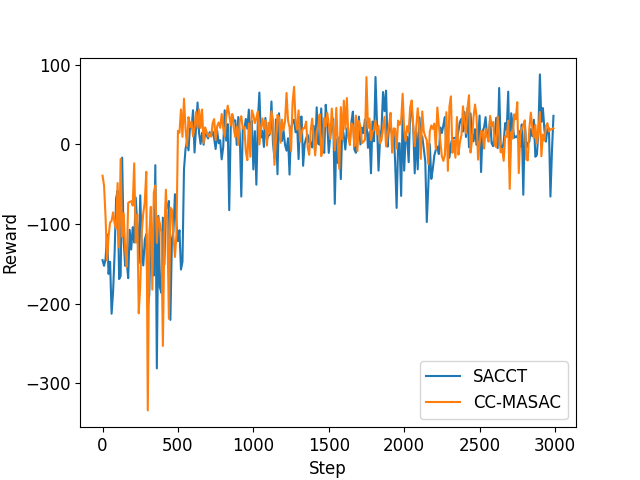}
%\label{fig:t_reward}
}
\newline
\subfloat[QoE]{\includegraphics[scale=0.48,clip]{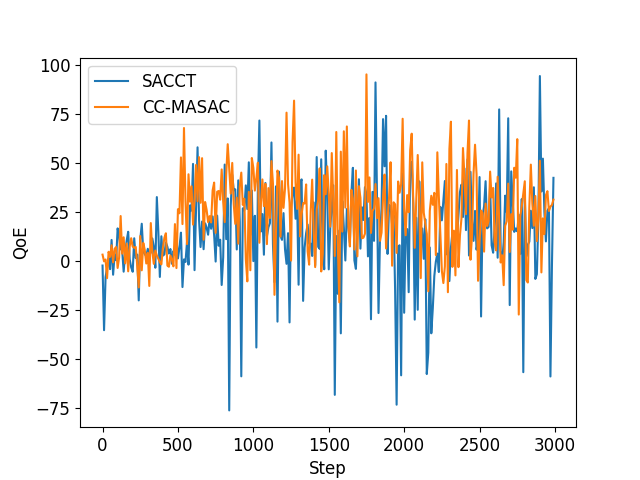}
%\label{fig:t_qoe}
}
\newline
\subfloat[Energy consumption]{\includegraphics[scale=0.48,clip]{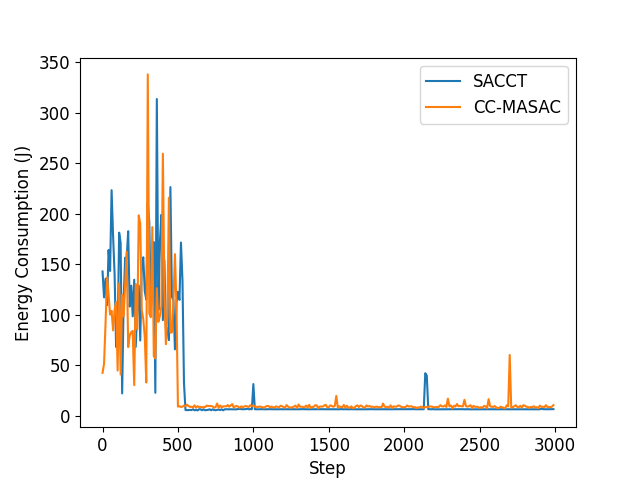}
%\label{fig:t_energy}
}
\caption{Convergence performance for SAC-based scheme.}
\label{fig:convergence}
\end{figure}

\figurename~\ref{fig:convergence} illustrates the convergence of the proposed \algname against SACCT, which is also a SAC-based scheme. The results indicate that our proposal can achieve higher and more stable expected rewards as the training proceeds. The reward herein is associated with QoE, energy consumption and task completion ratio as defined in Equation~\ref{eq:obj2}. Our proposal can enhance the QoE performance while maintaining a comparable energy performance, leading to an improved cumulative reward. Once the training process has converged after 3000 iterations, the average reward, QoE, and energy consumption for \algname are 14.78, 24.50, and 9.40 respectively. In comparison, for SACCT, these values are 7.03, 16.08, and 8.99.
The result highlights the significance of the MA-SAC mechanism, which involves multiple agents collaborating to learn from each other's experiences in order to achieve collaborative training to maximize the reward value. For each SAC-based scheme, we save the models obtained after 3000 steps for evaluating the learning policy.

\begin{figure*}[t]
	\centering
        \begin{minipage}[b]{0.9\columnwidth}
		\centering
		\includegraphics[width=\columnwidth]{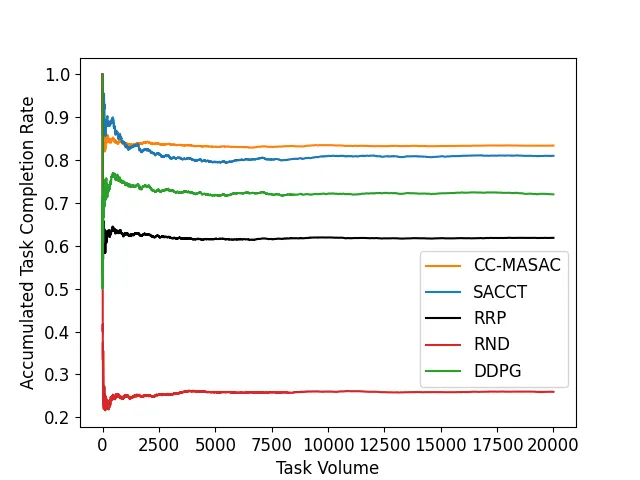}
		\subcaption{Task completion rate}\label{fig:tcr1}
	\end{minipage}	
	\begin{minipage}[b]{0.9\columnwidth}
		\centering
		\includegraphics[width=\columnwidth]{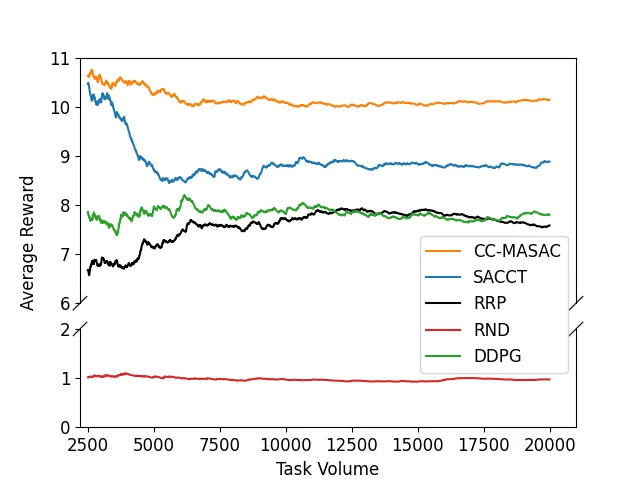}
		\subcaption{Average reward}\label{fig:qoe1}
	\end{minipage} \newline
	\begin{minipage}[b]{0.9\columnwidth}
		\centering
		\includegraphics[width=\columnwidth]{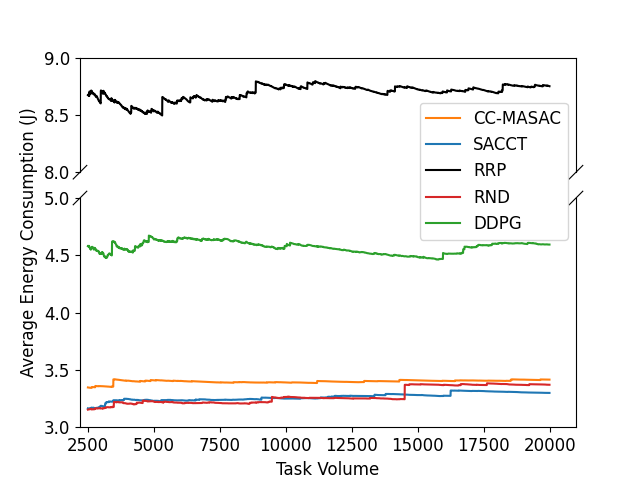}
		\subcaption{Average energy consumption}\label{fig:energy1}
	\end{minipage}
 	\begin{minipage}[b]{0.9\columnwidth}
		\centering
		\includegraphics[width=\columnwidth]{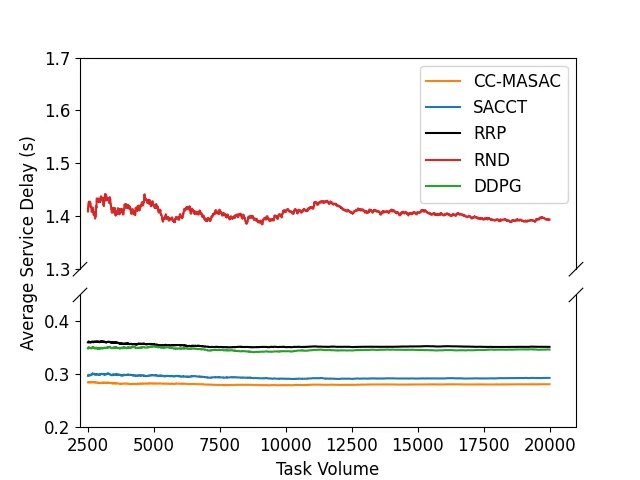}
		\subcaption{Average service delay}\label{fig:delay1}
	\end{minipage}
	\caption{Overall task offloading performance of different schemes.}
	\label{fig:overall_1}
\end{figure*}

Next, we evaluate the learning policy performance for different schemes in terms of various criteria. \figurename~\ref{fig:overall_1} shows the performance of each scheme across four criteria based on different computing task volumes in STN. Due to performance instability across the initial 2500 tasks, the average performance over these tasks was chosen as the starting point for plotting the performance graphs. The results in \figurename~\ref{fig:overall_1} demonstrate that our proposed \algname can achieve superior performance over other comparative schemes overall. In general, RND-MAXBR primarily focuses on balancing the utilization of heterogeneous resources within an STN, which can lead to the excessive use of resources from certain nodes. This limitation restricts the selection of resources for future computing tasks. Our proposed  \algname, aims to determine the optimal path selection through an analysis of factors such as service delay and resource utilization. Unlike SACCT, which employs a single agent, \algname collaborates with multiple agents to gather information from both the network and the tasks in order to train for optimal decision-making during the resource allocation and bitrate control process.

% The results in \figurename~\ref{fig:overall_1} demonstrate that our proposal can achieve superior performance over other comparative schemes in overall. Especially, \algname can significantly improve the task completion ratio performance against SACCT and RND-MAXBR by 
% 12\% and 33\% in average, respectively. The remaining three criteria QoE, energy consumption and service delay are calculated based on the successfully computed tasks. Similar to the trends observed during data training process (depicted in \figurename~\ref{fig:overall_1}), without sacrificing the performance of energy consumption, \algname can achieve superior QoE and service delay performance when handling the largest number of computing tasks: It surpasses the second-best scheme SACCT by a margin of 12\% in terms of QoE, with the enhancement in service delay increasing to 18\%. 

In detail, \algname improves the average task rewards compared to SACCT, DDPG, and RRP, with average increases of 16\%, 34\%, and 35\%, respectively. The SAC-based scheme (\algname and SAACCT) generally outperforms the DDPG-based one (DDPG) due to its utilization of stochastic policies. This feature enhances the exploration capabilities and robustness, leading to superior performance. Especially our proposed \algname incorporates not only self-observations and actions, but also those of all other agents, enabling more informed action selection based on current states and achieving higher rewards. Regarding task completion rates, after handling 1000 tasks, \algname achieves a completion rate of 83\%, surpassing SACCT (81\%), DDPG (72\%), RRP (61\%), and RND (25\%). Meanwhile, \algname also outperforms other comparative schemes in terms of service delay performance. From the energy consumption perspective, although RND and SACCT exhibit superior energy efficiency than \algname, the gap in energy consumption between \algname and these two algorithms is limited to an average of 3\%. Nevertheless, \algname significantly surpasses other schemes in terms of task completion rate, reward, and service delay, as aforementioned. In summary, our proposed \algname outperforms the other four comparative schemes in completion rate, reward, and service latency performance, while maintaining a low level of energy consumption.

%% file: 7_conclusion.tex
\section{Conclusion}~\label{sec:conclusion}
In this paper, we study the joint optimization problem for adaptive video streaming and task offloading in STNs. Since the original problem is NP-hard, we decompose the original optimization problem into two sub-problems to obtain sub-optimal results within a realistic calculation time. The first sub-problem, path selection, focuses on selecting an end-to-end path candidate that minimizes service delay while maintaining high resource utilization. The second sub-problem, resource allocation with adaptive bitrate control, concentrates on optimizing bitrate adaptation over the selected path to maximize QoE, task completion rate, and energy efficiency. Specifically, an MA-SAC based algorithm is developed to decide the optimal bitrate for the arriving video services based on the characteristics of video streaming and the resource usage of multiple nodes in STNs. Extensive simulations demonstrate the superiority of our proposal over baseline schemes in various criteria, including QoE, task completion rate, and service delay while maintaining low energy consumption level.%, and resource utilization efficiency.